\documentclass[12pt]{iopart}

\usepackage{iopams, cite}
\usepackage{graphicx}
\usepackage{bm}
\usepackage{color}

\usepackage[latin1]{inputenc}
\usepackage{bm}
\usepackage{multirow,amssymb,amsbsy,amsfonts,latexsym,fancyhdr,graphicx}
\makeatletter
\usepackage{pifont}

\newcommand{\Ignore}[1]{}

\definecolor{dgreen}{rgb}{0,0.5,0}

\definecolor{delete}{cmyk}{0.5,0,0,0}
\definecolor{deletey}{cmyk}{0,0.5,0,0}

\newcommand{\ket}[1]{\displaystyle{|#1\rangle}}
\newcommand{\bra}[1]{\displaystyle{\langle #1|}}
\newcommand{\ffi}{\varphi}
\newcommand{\sx}{\sigma_x}
\newcommand{\sy}{\sigma_y}
\newcommand{\sz}{\sigma_z}

\newcommand{\smn}{\sigma_-}

\begin{document}

\title{$GHZ$ state generation of three Josephson qubits in presence of bosonic baths}

\author{S. Spilla}
\address{Dipartimento di Fisica, Universit\`a di Palermo, via Archirafi 36,
90123 Palermo, Italy}

\author{R. Migliore}
\address{Institute of Biophysics, National Research Council, via U. La Malfa 153, 90146 Palermo, Italy}

\author{M. Scala}
\address{Dipartimento di Fisica, Universit\`a di Palermo, via Archirafi 36,
90123 Palermo, Italy}

\author{A. Napoli}
\address{Dipartimento di Fisica, Universit\`a di Palermo, via Archirafi 36,
90123 Palermo, Italy}

\date{\today}

\begin{abstract}
We analyze an entangling protocol to generate tripartite Greenberger-Horne-Zeilinger states in a system consisting of three superconducting qubits with pairwise coupling.
The dynamics of the open quantum system is investigated by taking into account  the interaction of each qubit with an independent bosonic bath with an ohmic spectral structure. To this end a microscopic master equation is constructed and exactly solved. We find that the protocol here discussed is stable against decoherence and dissipation due to the presence of the external baths.
\end{abstract}


\maketitle

\section{Introduction}
Since its introduction, quantum entanglement has played a central
role in foundational discussions of quantum mechanics. More
recently due to the advent of new more applicative areas, like
quantum information and communication fields, the concept of
entanglement has attracted a renewed interest from the scientific
community. Entangled quantum states have indeed proved to be
essential resources both for quantum information processing and
computational tasks. Also for this reason, in the last few years
many efforts have been devoted to the design and the
implementation, in very different physical areas, of schemes aimed
at generating entangled states \cite{Martinis, Migliore, E1, E2, E3, F, G,
H, cavita, cavita2, NMR, Ioni}. In this context, in particular,
superconductive qubits turned out to be promising candidates
providing their scalability and the possibility of controlling and
manipulating their quantum state in situ via external magnetic
field and voltage pulses \cite{Devoret, Levitov, Ploeg, Cosmelli}.

The efficiency of solid state architectures, however, is
unavoidably limited by decoherence and dissipation phenomena
related to the presence of different noise sources partly stemming
from control circuitry but also having microscopic origin. Thus,
having as final target the realization of states characterized by
prefixed quantum correlations, it is obviously important to
estimate the effects of the coupling between the system considered
and its surroundings.

Very recently Galiautdinov and Martinis \cite{Martinis} have
presented a protocol suitable for generating maximally entangled
states, namely $GHZ$ and $W$ states, of three Josephson qubits.
The key idea on which their proposal is based, is that for
implementing symmetric states, as the $GHZ$ and $W$ are, it is
convenient to symmetrically control all the qubits in the system.
In particular, making use of a triangular coupling interaction
scheme and exploiting single qubit local rotations, they
demonstrate the possibility of generating the desired state
appropriately setting the interaction time between the qubits. In
their analysis however, the authors considered the system as an
ideal one, without taking into account in any way its unavoidable
coupling with uncontrollable external degrees of freedom. In this
paper, following the idea proposed in ref.\cite{Martinis}, we
investigate on the effects of the environment on the generation of
$GHZ$ states. More in detail, we concentrate our attention on all
the external degrees of freedom that can be effectively modelized
as independent bosonic modes taking into account their presence
from the very beginning. We moreover exploit the same triangular
coupling mechanism envisaged in ref. \cite{Martinis} but we modify
the single qubit rotation protocols with respect the ones of
Galiautdinov and Martinis. Our analysis clearly prove that the
scheme for generating GHZ states is stable enough against
the noise sources we consider.

The paper is organized as follows: in section \ref{Entangling} we briefly
discuss the key ingredients of the Galiautdinov and Martinis (G-M)
procedure whereas in section \ref{rotations} we propose a possible way to
reduce the time required to generate the desired states. All the
steps of the generation protocol are then investigated in section \ref{MEDerivation}
supposing that each qubit of the system interacts with an
independent bosonic bath. The last section is devoted to the
discussion of the result we have obtained.

\section{Galiautdinov - Martinis entangling protocol} \label{Entangling}
In this section we briefly summarize the single step entangling
protocol, proposed by Galiautdinov and Martinis in order to
generate the three-qubits $GHZ$ states
\begin{equation}\label{GHZ}
\ket{GHZ}=\frac{1}{\sqrt{2}}(\ket{000}+e^{i\ffi}\ket{111})
\end{equation}
being $\ket{0}$ and $\ket{1}$ the ground and excited states of each qubit respectively.
More in particular we review some aspects of the procedure that are of interest in the context of the present paper.
The Hamiltonian model describing the physical system consisting of three Josephson qubits with pairwise coupling, is given by
\begin{equation}\label{HamCoupl}
H=\sum_{i=1}^3\Big(\frac{\omega}{2}\sz^{i}+\frac{1}{2}[g(\sx^{i}\sx^{i+1}+\sy^{i}\sy^{i+1})+\tilde{g}\sz^{i}\sz^{i+1}]\Big)
\end{equation}
with $\sigma_k^{4}=\sigma_k^1$ ($k=x,y,z$). Introducing the collective operators
\begin{equation}
\mathbf{S}=\sum_{j=1}^3\frac{\boldsymbol{\sigma}^j}{2}
\end{equation}
we can rewrite equation (\ref{HamCoupl}) in the following more convenient form
\begin{equation}\label{HamCouplS}
H=\omega S_z+gS^2-(g-\tilde{g})S_z^2
\end{equation}
within a constant term. Starting from equation (\ref{HamCouplS}) it is evident that the eigenstates of the system can be written as common eigenstates
$\ket{s_{12},s,m}$ of the operators $S_{12}^2=\bigr[\frac{1}{2}(\boldsymbol{\sigma}^1+\boldsymbol{\sigma}^2)\bigr]^2$, $S^2$ and $S_z$:
\begin{eqnarray}
\nonumber S_{12}^2\ket{s_{12},s,m}&=s_{12}(s_{12}+1)\ket{s_{12},s,m}\\
S^2\ket{s_{12},s,m}&=s(s+1)\ket{s_{12},s,m}\\
\nonumber S_z\ket{s_{12},s,m}&=m\ket{s_{12},s,m}
\end{eqnarray}
In particular it is immediate to convince oneself that the two states $\ket{000}\equiv\Bigl|s_{12}=1,s=\frac{3}{2},s_z=-\frac{3}{2}\Bigr\rangle$
and $\ket{111}\equiv\Bigl|s_{12}=1,s=\frac{3}{2},s_z=\frac{3}{2}\Bigr\rangle$ are eigenstates of $H$ correspondent to the eigenvalues
$-\frac{3}{2}\omega+\frac{3}{4}(2g+3\tilde{g})$ and $\frac{3}{2}\omega+\frac{3}{4}(2g+3\tilde{g})$ respectively.
In view of these considerations it is clear that, if at $t=0$ the three qubits are in their respective ground state,
in order to guide the system toward the desired state (\ref{GHZ}) it becomes necessary to implement some local rotations before
 turning on the interaction mechanism described by equation (\ref{HamCoupl}). This is what Galiautdinov and Martinis do,
 making thus sure an initial condition having both $\ket{000}$ and $\ket{111}$ components.
 The entanglement is then performed by switching on, for an appropriate interval of time $t_{int}=\frac{\pi}{2(g-\tilde{g})}$,
 the interaction described by equation (\ref{HamCoupl}) and finally by realizing an additional single-qubit rotation.
The scheme thus consists of three different steps: in the first
and the third ones, the Josephson qubits are independent and are
driven by external fields in order to appropriately rotate their
state. In the second step instead the three qubits are coupled
thus producing the desired entanglement among them.

\section{Single-qubit rotations}\label{rotations}
As we have underlined in section \ref{Entangling}, starting from
the initial condition $\ket{000}$ the interaction mechanism
described in equation (\ref{HamCoupl}) can be usefully exploited
for generating $GHZ$ states of three qubits, only if local
rotation operations are realized as first and final steps of the
procedure. These two distinct operations of the protocol require a
total time of realization $t_{13}=t_1+t_3$, which has to add to the
length of the qubit interaction time
$t_{int}=\frac{\pi}{2(g-\tilde{g})}$, if we wish to estimate the
total duration of the generation scheme. Thus the choice of the
physical mechanism able to perform appropriate single qubit
rotations, could be usefully exploited to control the time
required to generate the desired state starting from the state
$\ket{000}$. This aspect is of particular interest especially when
the presence of external degrees of freedom is not negligible. At
the light of these considerations we have chosen rotation
mechanisms different from the ones envisaged in
ref.\cite{Martinis}. In particular we suppose that in the first,
as well as in the last step of the scheme, whose duration is
hereafter indicated by $t_1$ and $t_3$ respectively, the system of
the three qubits is described by the following Hamiltonian
\begin{equation}\label{Hamrot3}
H_{rot}^l=\sum_{j=1}^3H_{rot}^l(j)\hspace{.8cm}(l=I,III)
\end{equation}
with
\begin{equation}\label{HamRot}
H_{rot}^l(j)=\frac{\omega}{2}\sz^j+\frac{\omega}{2}\left(e^{i\beta_l}\smn^j+\mathrm{h.c.}\right)
\end{equation}
where
\begin{eqnarray}
\beta_I&=\pi\left(\frac{1}{\sqrt{2}}+1\right)\\
\nonumber \beta_{III}&=\pi\left(\frac{3+\sqrt{2}}{2}\right)+\pi\left(\frac{(3+\sqrt{3})\omega}{8(g-\tilde{g})}\right).
\end{eqnarray}
In ref. \cite{Chirolli} is discussed in detail a possible way to
realize hamiltonian model like the one given by equation
(\ref{Hamrot3}) showing in particular that a full control of qubit
rotations on the entire Bloch sphere can be achieved.

It is possible to prove that setting
$t_1=t_3=\frac{\pi}{\sqrt{2}\omega}$, the sequence of the three
steps leads to the desired $GHZ$ states when the interaction
between the system of the three Josephson qubits and the external
world may be neglected. After some calculation indeed it is
possible to obtain that at $t=t_1$ the state of the system is
given by
\begin{eqnarray}\label{1rot}
\ket{\psi(t_1)}=\frac{1}{\sqrt{8}}\bigg(\ket{000}+\ket{111}e^{-\frac{3i\pi}{\sqrt{2}}}+\sqrt{3}\ket{W}e^{-\frac{i\pi}{\sqrt{2}}}
+\sqrt{3}\ket{W'}e^{-\sqrt{2}i\pi}\bigg).
\end{eqnarray}
where
\begin{equation}
\ket{W}=\frac{1}{\sqrt{3}}(\ket{100}+\ket{010}+\ket{001})
\end{equation}
and
\begin{equation}
\ket{W'}=\frac{1}{\sqrt{3}}(\ket{011}+\ket{101}+\ket{110}).
\end{equation}
At $t=t_1$ the interaction mechanism described by equation (\ref{HamCoupl}) is switched on for a time $t_{int}=\frac{\pi}{2(g-\tilde{g})}$.
At the end of this second step the state of the system will be
\begin{eqnarray}\label{2step}
\ket{\psi(t_1+t_{int})}=&\frac{1}{\sqrt{8}}\bigg(\ket{000}+\ket{111}e^{-\frac{3i\pi}{2}\big(\frac{\omega}{g-\tilde{g}}+\sqrt{2}\big)}+\\
\nonumber &-\sqrt{3}\ket{W}e^{i\pi(\frac{(\sqrt{3}-3)\omega}{4(g-\tilde{g})}-\frac{\sqrt{2}}{2})}
-\sqrt{3}\ket{W'}e^{-i\pi(\frac{(\sqrt{3}+3)\omega}{4(g-\tilde{g})}+\sqrt{2})}\bigg).
\end{eqnarray}
Thus the last step of the procedure described by $H_{rot}^{III}$, leads the system into the final state
\begin{equation}\label{StatoFinale}
\ket{\psi(t_1+t_{int}+t_3)}=\frac{1}{2}\bigg[(i+e^{i\alpha})\ket{000}+ie^{-i\theta}(i-e^{i\alpha})\ket{111}\bigg]
\end{equation}
with
\begin{eqnarray}\label{alfateta}
\nonumber &\alpha=\frac{3\pi\omega(\sqrt{3}-1)}{8(g-\tilde{g})}\\
&\theta=\frac{3\pi\sqrt{2}}{2}+\frac{3\pi\omega(\sqrt{3}+3)}{8(g-\tilde{g})}.
\end{eqnarray}
Starting from equations (\ref{StatoFinale}) and (\ref{alfateta}) it is immediate to convince oneself that, if the condition
\begin{equation}\label{CondGHZ}
\frac{\omega}{g-\tilde{g}}=\frac{8k}{3(\sqrt{3}-1)}\hspace{.8cm}k\in\mathbb{N}
\end{equation}
is satisfied, the three Josephson qubits are left in the desired $GHZ$ state.\\
Thus we can say that the time required to generate the state (\ref{GHZ}) starting from the condition $\ket{000}$ can be estimated as
\begin{equation}
t_{tot}=t_1+t_{int}+t_3=\frac{\sqrt{2}\pi}{\omega}+\frac{\pi}{2(g-\tilde{g})}.
\end{equation}
This value of $t_{tot}$ has to be compared with a time
$t_{tot}=\frac{\pi}{\Omega}+\frac{\pi}{2(g-\tilde{g})}$, with
$\Omega\ll\omega$ required if the procedure of Galiautdinov and
Martinis is adopted. Starting indeed from the results presented in
refs \cite{Martinis} and \cite{Geller}, it is possible to convince
oneself that the proposal of Galiautdinov and Martinis requires
that in the first an in the last step of the procedure the
dynamics of each qubit is governed by the following Hamiltonian
model

\begin{equation}\label{GM}
H_{GM}^{I,III}=\sum_j H_{GM}^{I,III}(j)
\end{equation}
where
\begin{equation}\label{GMj}
H_{GM}^{I,III}(j)=\frac{\omega}{2}\sigma_z^j+\Omega cos(\omega
t+\phi_{I,III})\sigma_x^j
\end{equation}
with $\Omega\ll \omega$ and $\phi_I=-\frac{\pi}{2}$
$\phi_{III}=0$.

Thus, changing the way to rotate the state of the three
qubits during the first and the last step of the procedure, is possible
to reduce the time required to generate
the target state. As said before, this aspect is of particular
relevance when the interaction of the system with the external
world is not negligible. The price to pay anyway is that in our
case, differently from the scheme of Galiautdinov and Martinis,
three qubit GHZ states can be generated only if the condition given in
equation (\ref{CondGHZ}) is satisfied. Generally speaking, indeed, at the
end of the procedure, the three Josephson qubits are left in a
linear superposition of the two states $\ket{000}$ and $\ket{111}$
with amplitudes $A_{000}=\frac{1}{2}(i+e^{i\alpha})$ and
$A_{111}=\frac{i}{2}e^{-i\theta}(i-e^{i\alpha})$ respectively. It
is important however, to stress that the condition (\ref{CondGHZ})
is compatible with typical values of the free frequency $\omega$,
that generally speaking can be taken of the order of 10GHz, and
with the values of the coupling constants $g$ and $\tilde{g}$,
that reasonably can be assumed of the order of 1GHz and
$10^{-1}$GHz respectively \cite{You, Makhlin, DevoMartin,ScalaMigliore}. On the other hand, condition
(\ref{CondGHZ}) is not so mandatory as it appears, since we have verified
that variations of ten percent in the ratio
$\frac{\omega}{g-\tilde{g}}$ are still compatible
 with the requirement that $|A_{000}|^2\simeq|A_{111}|^2$.

\section{Microscopic master equation derivation}\label{MEDerivation}
In a realistic description of the scheme until now discussed we
cannot neglect the presence of uncontrollable external degrees of
freedom coupled to the three Josephson qubits that, generally
speaking, affects in a bad way quantum state generation protocols.
These degrees of freedom, that define the so-called environment,
can have different physical origin and thus different
descriptions. In this section we will focus our attention on all
the external degrees of freedom describable as independent bosonic
modes \cite{bos1, bos2, bos3, bos4, bos5, bos6, bos7, bos8, bos9}.
More in detail, we will suppose that during all the process
each qubit is coupled to a bosonic bath and the three baths are
independent.

The plausibility of this assumption can be tracked back to the
fact that the three superconductive qubits are spatially separated
so that it is reasonable to suppose that each of them is affected
by sources of noise stemming from different parts of the
superconductive circuit. In this section we review all the three
steps of the procedure before discussed, analyzing the dynamics of
the system by considering from the very beginning the interaction
of each qubit with a bosonic bath. In order to do this we will
construct and solve microscopic master equations in correspondence
to the three different steps described in section \ref{rotations} in which
the generation scheme is structured. In each of the three steps the master
equation will be derived in the Born - Markov and Rotating wave approximations
\cite{Petruccione}. We wish to stress at this point that the use of
microscopic master equations instead of naiver and more popular
phenomenological ones, becomes important particularly when
structured reservoirs are considered \cite{Migliore_2011}.

\subsection{First step: single-qubit rotation}

Let us suppose that the three Josephson qubits are initially
prepared in the ground state $\ket{000}$ and that the Hamiltonian
describing the system in the first step of the procedure is given
by equation (\ref{Hamrot3}) with $l=I$. Each qubit moreover is coupled
to a bosonic bath and the three baths are independent. The
Hamiltonian model describing the system in the first step can be
thus written as \cite{Scala}
\begin{equation}\label{HamRes}
H_1=H_{rot}^I+H_B+H_{int}
\\\end{equation}
with
\begin{eqnarray}\label{HamRes1}
H_B\equiv&H_B(1)+H_B(2)+H_B(3)=\\
\nonumber =&\sum_k\omega_k^1a_k^{1\dag} a_k^1+\sum_k\omega_k^2a_k^{2\dag} a_k^2+\sum_k\omega_k^3a_k^{3\dag} a_k^3
\end{eqnarray}
and
\begin{eqnarray}\label{HamRes2}
H_{int}=\sx^1\otimes\sum_kg_k^1(a_k^1+& a_k^{1\dag})+\sx^2\otimes\sum_kg_k^2(a_k^2+ a_k^{2\dag})+\\ \nonumber&+\sx^3\otimes\sum_kg_k^3(a_k^3+ a_k^{3\dag}).
\end{eqnarray}
Exploiting standard procedure \cite{Petruccione} we now derive the microscopic
master equation suitable to describe the dynamics of the three
qubits system. Taking into account the fact that the qubits, as
well as the baths, are, in this case, independent, it is enough
to construct and solve the master equation
correspondent to a single superconductive qubit. Indicating by
$\rho_j(t)$ the density matrix of the $j$-th $(j=1,2,3)$ qubit, it
is possible to prove that during the first step we have
\begin{eqnarray}\label{MasterRot1}
\nonumber &\dot{\rho}_j(t)=-i[H_{rot}^I(j),\rho_j(t)]+\\&\gamma_1(\omega_1)\biggr(A_j(\omega_1)\rho_j(t)A_j^\dag(\omega_1)-
\frac{1}{2}\big\{A_j^\dag(\omega_1)A_j(\omega_1),\rho_j(t)\big\}\biggr)+\\
\nonumber &\gamma_1(\omega_{2})\biggr(A_j(\omega_{2})\rho_j(t)A_j^\dag(\omega_{2})-
\frac{1}{2}\big\{A_j^\dag(\omega_{2})A_j(\omega_{2}),\rho_j(t)\big\}\biggr)
\end{eqnarray}
where the Bohr frequencies are respectively
$\omega_1=\sqrt{2}\omega$ and $\omega_2=0$ whereas the
correspondent operators, describing the jumps between the
eigenstates $\ket{\psi_\pm\epsilon}_j$
($\epsilon=\frac{\omega}{\sqrt{2}}$), of the Hamiltonian
$H_{rot}^I(j)$, are given by
\begin{eqnarray}\label{A1fase}
\nonumber A_j(\omega_1)&\equiv\ket{\psi_{-\epsilon}}_{jj}\bra{\psi_{-\epsilon}}\sigma_x^j\ket{\psi_\epsilon}_{jj}\bra{\psi_\epsilon}=\\
\nonumber &=\bigg(\frac{1}{\sqrt{2}}\cos{\frac{\pi}{\sqrt{2}}}-i\sin{\frac{\pi}{\sqrt{2}}}\bigg)\ket{\psi_{-\epsilon}}_{jj}\bra{\psi_{\epsilon}}\\
A_j(\omega_{2})&\equiv\ket{\psi_{-\epsilon}}_{jj}\bra{\psi_{-\epsilon}}\sigma_x^j\ket{\psi_{-\epsilon}}_{jj}\bra{\psi_{-\epsilon}}+\\
\nonumber &+\ket{\psi_\epsilon}_{jj}\bra{\psi_\epsilon}\sigma_x^j\ket{\psi_\epsilon}_{jj}\bra{\psi_\epsilon}=\\
\nonumber &=\frac{1}{\sqrt{2}}\cos{\frac{\pi}{\sqrt{2}}}\bigg(\ket{\psi_{-\epsilon}}_{jj}\bra{\psi_{-\epsilon}}-\ket{\psi_\epsilon}_{jj}\bra{\psi_\epsilon}\bigg)
\end{eqnarray}
with
\begin{eqnarray}\label{AutovRot1}
&\ket{\psi_\epsilon}_j=\frac{1}{2}\left[\sqrt{2+\sqrt{2}}e^{-i\beta_I}\ket{1}_j+ \sqrt{2-\sqrt{2}}\ket{0}_j\right]\\
\nonumber &\ket{\psi_{-\epsilon}}_j=\frac{1}{2}\left[\sqrt{2-\sqrt{2}}e^{-i\beta_I}\ket{1}_j- \sqrt{2+\sqrt{2}}\ket{0}_j\right].
\end{eqnarray}
Concerning the decay rates $\gamma_1(\omega_1)$ and
$\gamma_1(\omega_2)$ appearing in equation (\ref{MasterRot1}),
we will fix their numerical value in the next section where we explicitly
give the spectral properties of the baths.

\subsection{Second step: entangling procedure}
As we have previously discussed, the next step requires that the
three qubits interact among them through the coupling mechanism
described by equation (\ref{HamCouplS}). In addition each qubit
interacts with a bosonic bath. Thus the Hamiltonian describing the
total system in this second step can be written as
\begin{equation}
H_2=H+H_B+H_{int}.
\end{equation}
The master equation for the
density matrix of the three qubits during the second step can be written in the form
\begin{eqnarray}\label{MasterCoupl}
\dot{\rho}(t)=&-i[H,\rho(t)]+\\
\nonumber &+\sum_{j=1}^3\sum_{k=1}^8\gamma_2(\omega_k)\biggr(A_j(\omega_k)\rho(t)A_j^\dag(\omega_k)-\frac{1}{2}\big\{A_j^\dag(\omega_k)A_j(\omega_k),\rho(t)\big\}\biggr)
\end{eqnarray}
where the Bohr frequencies are the following
\begin{eqnarray}\label{BohrFr2}
\nonumber \omega_{1,3}&=\frac{3-\sqrt{3}}{2}\omega\pm2(g-\tilde{g})\\
\nonumber \omega_{2,4}&=\frac{3}{2}\omega\mp(g+2\tilde{g})\\
\omega_5&=\sqrt{3}\omega\\
\nonumber \omega_{6,7}&=\frac{\sqrt{3}}{2}\omega\pm3g\\
\nonumber \omega_8&=0
\end{eqnarray}
whereas the jump operators between the eigenstates of the
Hamiltonian (\ref{HamCouplS}) are given for convenience in
appendix B. We wish to underline that equation (\ref{MasterCoupl})
does not contain mixed terms of the form
$A_j(\omega_k)\rho(t)A_{j'}^\dag(\omega_k)-
\frac{1}{2}\big\{A_j^\dag(\omega_k)A_{j'}(\omega_k),\rho(t)\big\}$
with $j\neq j'$ in view of the fact that the three bosonic baths
are independent.

We have solved the master equation (\ref{MasterCoupl}),
considering as initial condition the solution of the master
equation (\ref{MasterRot1}) obtained in the previous paragraph at
$t=t_1$ . More in detail, taking into account the fact that the
ideal scheme provides a dynamics confined in the subspace
generated by the states $\ket{000}$, $\ket{111}$, $\ket{W}$ and
$\ket{W'}$, we have focused our attention on the projection of
$\rho(t)$ on this subspace. It is possible indeed to prove that
the neglected subspace will be at the most populated with a
probability not exceeding the 3\%.

\subsection{Third step: local rotations}
To complete the analysis of the GHZ state generation procedure in
presence of noise, we have to construct the microscopic master
equation describing the system in the last step of the scheme.
Actually it can be immediately deduced from the master equation
derived in the first step simply substituting $\beta_I$ with
$\beta_{III}$  in the eigenstates $\ket{\psi_{\pm\epsilon}}$
appearing in the jump operators. However, in this case it is more
convenient to write the jump operators exploiting the basis
\begin{eqnarray}\label{baseRot}
\nonumber &\ket{\widetilde{000}}=T\ket{000}\\
\nonumber &\ket{\widetilde{111}}=T\ket{111}\\
\nonumber &\ket{\widetilde{W}}=T\ket{W}=\frac{1}{\sqrt{3}}T(\ket{100}+\ket{010}+\ket{001})\\
\nonumber &\ket{\widetilde{W'}}=T\ket{W'}=\frac{1}{\sqrt{3}}T(\ket{011}+\ket{101}+\ket{110})\\
&\ket{\widetilde{\psi_1}}=T\ket{\psi_1}=\frac{1}{\sqrt{2}}T(\ket{100}-\ket{010})\\
\nonumber &\ket{\widetilde{\psi_1'}}=T\ket{\psi_1'}=\frac{1}{\sqrt{2}}T(\ket{011}-\ket{101})\\
\nonumber &\ket{\widetilde{\psi_2}}=T\ket{\psi_2}=\frac{1}{\sqrt{6}}T(\ket{100}+\ket{010}-2\ket{001})\\
\nonumber &\ket{\widetilde{\psi_2'}}=T\ket{\psi_2'}=\frac{1}{\sqrt{6}}T(\ket{011}+\ket{101}-2\ket{110}).
\end{eqnarray}
instead of the standard one. In this new basis the unitary
operator $T$ can be represented as
\begin{eqnarray}\label{TTrasf}
T=\frac{\sqrt{2+\sqrt{2}}}{8}\left(\begin{array}{ccc}
B_1 & & 0 \\ & B_2 & \\ 0 & & B_3 \\
\end{array}\right)
\end{eqnarray}
with
\begin{eqnarray}
\nonumber B_1=&{\scriptsize\left(\begin{array}{cccc}
 -(2+\sqrt{2}) & -(4-3\sqrt{2}) & \sqrt{6} & -\sqrt{3}(2-\sqrt{2})\\
 -(4-3\sqrt{2})e^{-3i\beta_{III}} & (2+\sqrt{2})e^{-3i\beta_{III}} &  \sqrt{3}(2-\sqrt{2})e^{-3i\beta_{III}} & \sqrt{6}e^{-3i\beta_{III}} \\
 \sqrt{6}e^{-i\beta_{III}} & \sqrt{3}(2-\sqrt{2})e^{-i\beta_{III}} & (3\sqrt{2}-2)e^{-i\beta_{III}} & -(4-\sqrt{2})e^{-i\beta_{III}} \\
 -\sqrt{3}(2-\sqrt{2})e^{-2i\beta_{III}} & \sqrt{6}e^{-2i\beta_{III}} &  -(4-\sqrt{2})e^{-2i\beta_{III}} & -(3\sqrt{2}-2)e^{-2i\beta_{III}} \\
 \end{array}\right)}\\
\nonumber B_2=&\left(\begin{array}{cc}
4e^{-i\beta_{III}} & 4(\sqrt{2}-1)e^{-i\beta_{III}} \\ 4(\sqrt{2}-1)e^{-2i\beta_{III}} & -4e^{-2i\beta_{III}} \\
\end{array}\right)\\
\nonumber B_3=&\left(\begin{array}{cc}
4e^{-i\beta_{III}} & 4(\sqrt{2}-1)e^{-i\beta_{III}} \\ 4(\sqrt{2}-1)e^{-2i\beta_{III}} & -4e^{-2i\beta_{III}} \\
\end{array}\right)
\end{eqnarray}
It is possible to demonstrate that in this case the master equation can be written as
\begin{eqnarray}\label{MasterE2Rot}
\dot{\rho}(t)=&-i[H_{rot}^{III},\rho(t)]+\\
\nonumber &+\sum_{j=1}^3\sum_{i=1}^{2}\gamma_3(\omega_i)\biggr(A_j(\omega_i)\rho(t)A_j^\dag(\omega_i)-\frac{1}{2}\big\{A_j^\dag(\omega_i)A_j(\omega_i),\rho(t)\big\}\biggr)
\end{eqnarray}
where the Bohr frequencies are the same as the ones in the first
step whereas the jump operators $A_j(\omega_i)$ between the
eigenstates of the Hamiltonian $H_{rot}^{III}$ are for convenience
given in the appendix C.

\section{Results and conclusions} \label{Conclu}

Having at disposal the microscopic master equations
(\ref{MasterRot1}), (\ref{MasterCoupl}) and (\ref{MasterE2Rot}),
describing the dynamics of the three-qubit system, we have found
the density matrix $\rho(t_{tot})$ of the system at the time instant
$t_{tot}=t_1+t_{int}+t_3$, supposing that at $t=0$ the initial
condition was $\ket{\psi(0)}=\ket{000}$. Moreover, we have assumed that all the three
baths were characterized by the same spectral density given in
particular by the ohmic one
\begin{equation}\label{ohmicspectrum}
\gamma(\omega)=\Bigg \{
 \begin{array}{c}
  \gamma_0 \;\;\;\;\; \omega=0\\
  \alpha \omega \;\;\;\;\; \omega\neq 0
\end{array}\end{equation}
where $\gamma_0$ is introduced in order to take into account a non
zero decay rate for $\omega=0$.

\begin{figure}
\begin{center}
\includegraphics[width=0.8\linewidth]{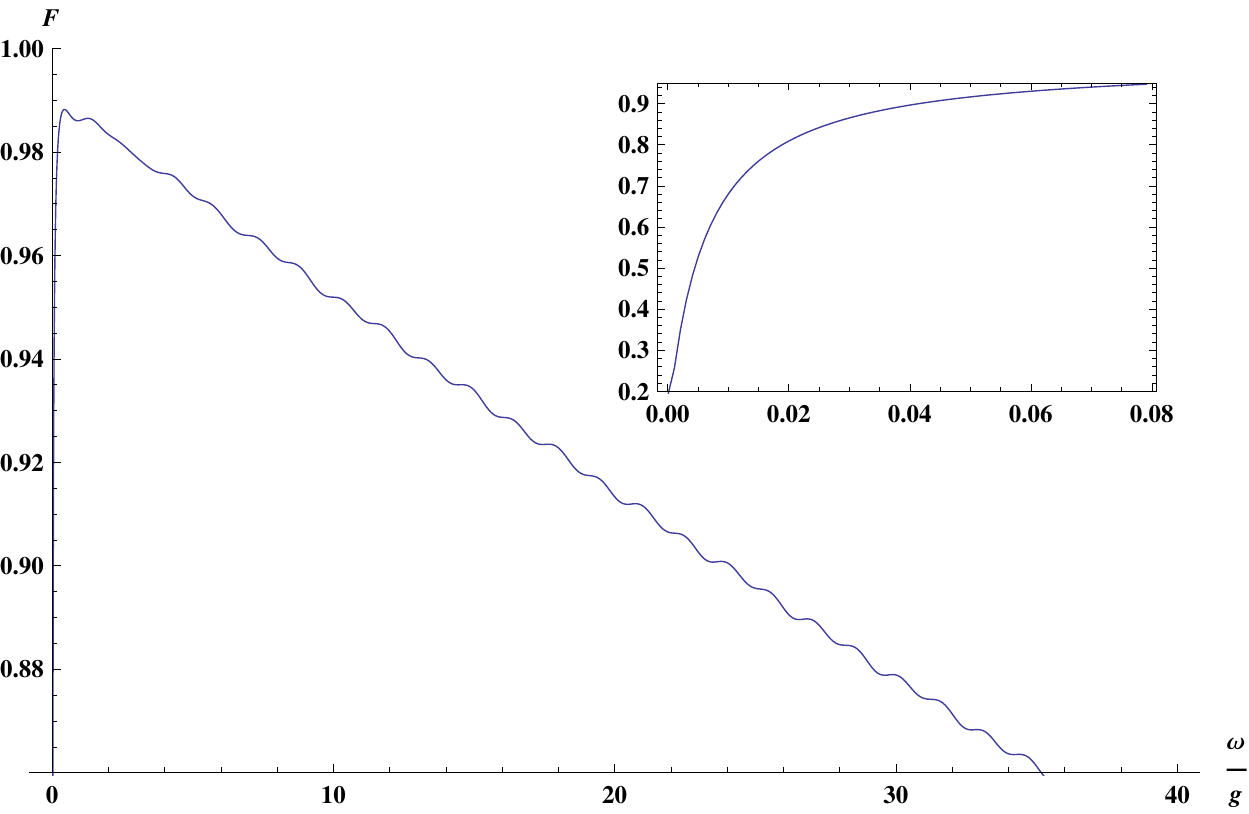}
\caption{ Fidelity $F$ as a function of $\omega/g$ when we assume that all the three baths were characterized by the same ohmic spectral density given in (\ref{ohmicspectrum}) with  $\gamma_0=\alpha=10^{-3}$ and the parameters assume the values $\tilde{g}/g=0.1$. The inset shows the Fidelity for values $0<\omega/g<0.08$.} \label{fig1}
\end{center}
\end{figure}

To quantify the effects of the bosonic baths we can consider the
fidelity $\emph{F}$

\begin{equation}\label{fidelity1}
\emph{F}=Tr\{\rho_{exp}\rho(t_{tot})\}
\end{equation}

that gives an idea of the difference existing between the density matrix $\rho_{exp}$, obtained when the interaction with the three baths is neglected, and the density matrix $\rho(t_{tot})$. The results we
have obtained are for convenience given in figure \ref{fig1} where we
plot \emph{F} as a function of the ratio $\omega/g$
assigning to the parameters $\gamma_0$ and $\alpha$ physically
reasonable values. In particular we have chosen
$\gamma_0=\alpha=10^{-3}$ \cite{Makhlin, You, DevoMartin, ScalaMigliore} .

\begin{figure}
\begin{center}
\includegraphics[width=0.8\linewidth]{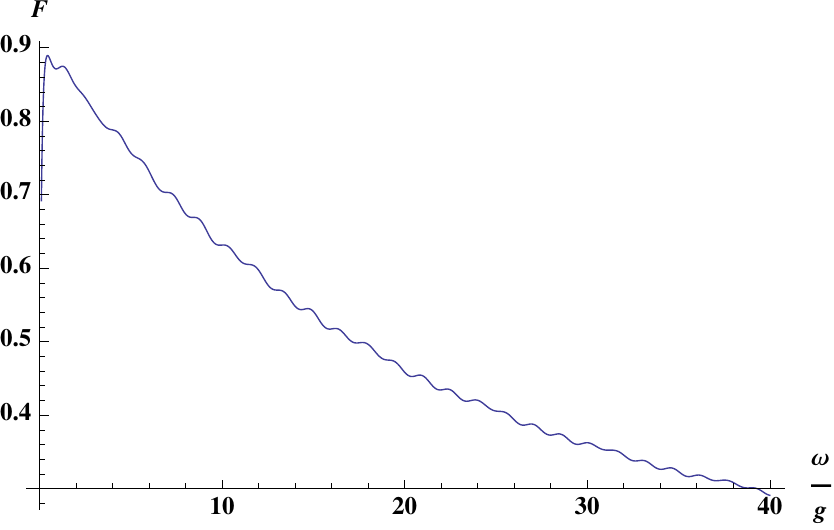}
\caption{Fidelity $F$ as a function of $\omega/g$ when we assume that all the three baths were characterized by the same ohmic spectral density given in (\ref{ohmicspectrum}) with  $\gamma_0=\alpha=10^{-2}$ and the parameters assume the values $\tilde{g}/g=0.1$.} \label{fig2}
\end{center}
\end{figure}

As we can see, at least for $\omega\lesssim 20g$ the presence of
bosonic baths at zero temperature does not affect in a
significative way the dynamics of the system during the different
steps of the procedure, being the fidelity not less that 0.90.
One should expect that the fidelity $F$ is a monotonically decreasing
function of $\omega/g$. The model we have used for the decay rate (see
eq.(\ref{ohmicspectrum}), however, is discontinuous  for zero frequency
because we want to consider also possible dephasing channels. In view of
this discontinuity one is not allowed to perform the limit $\omega/g$
tending to zero in the fidelity. Anyway this is not a problem in view of
the fact that for $\omega=0$ our scheme is meaningless since in this limit
no rotations are performed. Moreover, as the inset in figure 1 shows, the
increase of $F$ is rapid with respect to $\omega/g$.\\
Let us now observe that increasing of an order of magnitude the bath decay rates, the
fidelity $F$ remains experimentally significative as shown in
figure \ref{fig2}.

Both figures make evident that the presence of the three
independent bosonic baths does not affect in a dramatic way the
results reached under the hypothesis of perfect isolation.

\begin{figure}
\begin{center}
\includegraphics[width=0.8\linewidth]{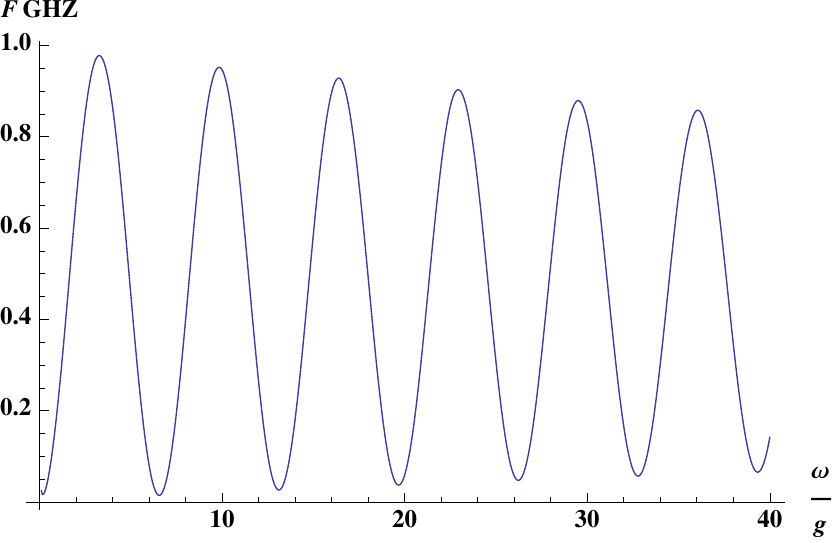}
\caption{Fidelity $F_{GHZ}$ as a function of $\omega/g$ when we assume that all the three baths were characterized by the same ohmic spectral density given in (\ref{ohmicspectrum}) with  $\gamma_0=\alpha=10^{-3}$ and the parameters assume the values $\tilde{g}/g=0.1$. To evaluate $F_{GHZ}$ we put $\ffi=\frac{3\pi}{2}(\sqrt{3}+\frac{(3+\sqrt{3})}{4}\frac{\omega}{g-\tilde{g}})$ for the GHZ state in equation (\ref{fidelity2}).} \label{fig3}
\end{center}
\end{figure}

\begin{figure}
\begin{center}
\includegraphics[]{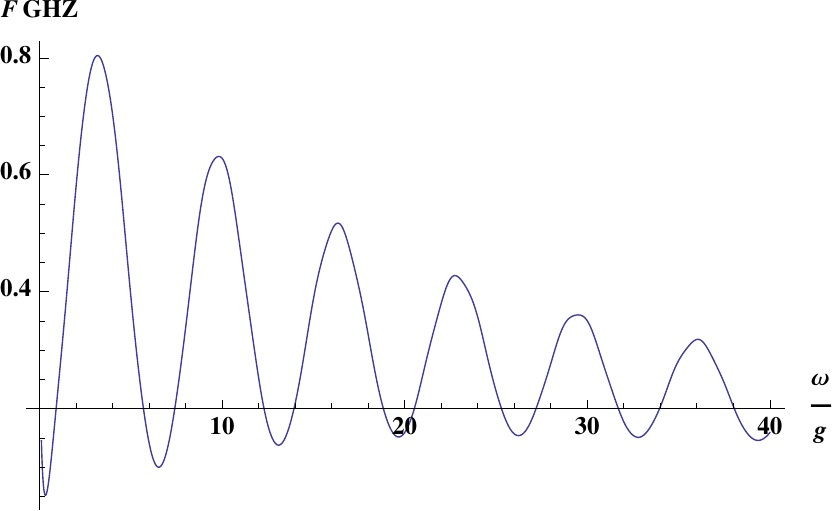}
\caption{Fidelity $F_{GHZ}$ as a function of $\omega/g$ when we assume that all the three baths were characterized by the same ohmic spectral density given in (\ref{ohmicspectrum}) with  $\gamma_0=\alpha=10^{-2}$ and the parameters assume the values $\tilde{g}/g=0.1$. To evaluate $F_{GHZ}$ we put $\ffi=\frac{3\pi}{2}(\sqrt{3}+\frac{(3+\sqrt{3})}{4}\frac{\omega}{g-\tilde{g}})$ for the GHZ state in equation (\ref{fidelity2}).} \label{fig4}
\end{center}
\end{figure}

We are however interested to the generation of GHZ states as given
in equation (\ref{GHZ}), which, as we have previously seen, can be
obtained only if the condition (\ref{CondGHZ}) is satisfied. Thus
it is of interest for us to analyze the fidelity $\emph{F}_{GHZ}$
defined as
\begin{equation}\label{fidelity2}
\emph{F}_{GHZ}=Tr\{\ket{GHZ}\bra{GHZ}\rho(t_{tot})\}
\end{equation}
and reported in figures \ref{fig3} and \ref{fig4} as a function of the ratio
$\omega/g$.

Figure \ref{fig3} is obtained in correspondence of bath decay rates
generally reported in literature as realistic ones \cite{Makhlin, You, DevoMartin, ScalaMigliore}
whereas the results reported in figure \ref{fig4} are obtained supposing
worse conditions. As expected, the fidelity $\emph{F}_{GHZ}$ shows
maxima in correspondence to values of $\omega/g$ such to
satisfy condition (\ref{CondGHZ}). The value of such maxima
moreover decreases increasing the ratio $\omega/g$. This
circumstance is in turn related to the fact that the decay rates
appearing in the master equations (\ref{MasterRot1}),
(\ref{MasterCoupl}) and (\ref{MasterE2Rot}), are increasing
functions of $\omega$. However, also considering the worst case we
may conclude that it is possible to choose an interval of values
of the ratio $\omega/g$ in correspondence of which
$\emph{F}_{GHZ}$ is greater than 0.7. On the other hand for
experimentally reasonable values of the decay rates $\gamma_0$ and
$\alpha$ we can obtain values of $\emph{F}_{GHZ}$ greater than
0.9 also fixing $\omega/g$ in different intervals, see figure \ref{fig3}.
Notwithstanding these values of the fidelity are less than the fault
tolerance threshold, they  are surely of interest in the context of
generation schemes of quantum states having assigned properties.

We thus may conclude that the scheme before discussed to generate
GHZ states (\ref{GHZ}) is robust enough with respect to the
presence of noise sources describable as independent bosonic
baths.

\section*{acknowledgments}
The Authors thank the Project PRIN 2008C3JE43\_003 for financial
support.

\section{Appendix A}

The eigenstates of the Hamiltonian given in (\ref{HamCouplS}) can
be written as common eigenstates $\ket{s_{12},s,m}$ of the
operators
$S_{12}^2=\bigr[\frac{1}{2}(\boldsymbol{\sigma}^1+\boldsymbol{\sigma}^2)\bigr]^2$,
$S^2$ and $S_z$ and can be cast in the form
\begin{eqnarray}\label{base}\
\nonumber &\Bigl|1,\frac{3}{2},-\frac{3}{2}\Bigr\rangle=\ket{000}\\
\nonumber &\Bigl|1,\frac{3}{2},\frac{3}{2}\Bigr\rangle=\ket{111}\\
\nonumber &\Bigl|1,\frac{3}{2},-\frac{1}{2}\Bigr\rangle=\frac{1}{\sqrt{3}}(\ket{100}+\ket{010}+\ket{001})\equiv\ket{W}\\
&\Bigl|1,\frac{3}{2},\frac{1}{2}\Bigr\rangle=\frac{1}{\sqrt{3}}(\ket{011}+\ket{101}+\ket{110})\equiv\ket{W'}\\
\nonumber &\Bigl|0,\frac{1}{2},-\frac{1}{2}\Bigr\rangle=\frac{1}{\sqrt{2}}(\ket{100}-\ket{010})\equiv=\ket{\psi_1}\\
\nonumber &\Bigl|0,\frac{1}{2},\frac{1}{2}\Bigr\rangle=\frac{1}{\sqrt{2}}(\ket{011}-\ket{101})\equiv\ket{\psi_1'}\\
\nonumber &\Bigl|1,\frac{1}{2},-\frac{1}{2}\Bigr\rangle=\frac{1}{\sqrt{6}}(\ket{100}+\ket{010}-2\ket{001})\equiv\ket{\psi_2}\\
\nonumber &\Bigl|1,\frac{1}{2},\frac{1}{2}\Bigr\rangle=\frac{1}{\sqrt{6}}(\ket{011}+\ket{101}-2\ket{110})\equiv\ket{\psi_2'}.
\end{eqnarray}
The correspondent eigenvalues are given by
\begin{eqnarray}\label{A.valoriCoupl}
\nonumber &E_{\ket{000}}=-\frac{3}{2}(\omega-\tilde{g})\\
\nonumber &E_{\ket{111}}=\frac{3}{2}(\omega+\tilde{g})\\
&E_{\ket{W}}=-\frac{1}{2}(\sqrt{3}\omega-4g+\tilde{g})\\
\nonumber &E_{\ket{W'}}=\frac{1}{2}(\sqrt{3}\omega+4g-\tilde{g})\\
\nonumber &E_{\ket{\psi_r}}=-(g+\frac{\tilde{g}}{2})\;\;\;\; \mathrm{for}\;\;\;
r=1,1',2,2'
\end{eqnarray}

\section{Appendix B}

The Bohr frequencies of the system in the second step are given in
(\ref{BohrFr2}) and the correspondent jump operators are
respectively

\begin{eqnarray}
\nonumber A_j(\omega_1)=&\frac{1}{\sqrt{3}}\ket{000}\bra{W}\;\;\;\; j=1,2,3\\
\nonumber A_j(\omega_2)=&(-1)^{j+1}\frac{1}{\sqrt{2}}\ket{000}\bra{\psi_1}+\frac{1}{\sqrt{6}}\ket{000}\bra{\psi_2}\;\;\;j=1,2\\
\nonumber A_3(\omega_2)=&-\sqrt{\frac{2}{3}}\ket{000}\bra{\psi_2}\\
\nonumber A_j(\omega_3)=&\frac{1}{\sqrt{3}}\ket{W'}\bra{111}\;\;\;j=1,2,3\\
\nonumber A_j(\omega_4)=&(-1)^{j+1}\frac{1}{\sqrt{2}}\ket{\psi_1'}\bra{111}+\frac{1}{\sqrt{6}}\ket{\psi_2'}\bra{111}\;\;\;j=1,2\\
\nonumber A_3(\omega_4)=&-\sqrt{\frac{2}{3}}\ket{\psi_2'}\bra{111}\\
\nonumber A_j(\omega_5)=&\frac{2}{3}\ket{W}\bra{W'}\;\;\;j=1,2,3\\
A_j(\omega_6)=&(-1)^j\frac{1}{\sqrt{6}}\ket{W}\bra{\psi_1'}-\frac{1}{3\sqrt{2}}\ket{W}\bra{\psi_2'}\;\;\;j=1,2\\
\nonumber A_3(\omega_6)=&\frac{\sqrt{2}}{3}\ket{W}\bra{\psi_2'}\\
\nonumber A_j(\omega_7)=&-\frac{1}{\sqrt{6}}\ket{\psi_1}\bra{W'}-\frac{1}{3\sqrt{2}}\ket{\psi_2}\bra{W'}\;\;\;j=1,2\\
\nonumber A_3(\omega_7)=&\frac{\sqrt{2}}{3}\ket{\psi_2}\bra{W'}\\
\nonumber A_j(\omega_8)=&(-1)^{j+1}\frac{1}{\sqrt{3}}(\ket{\psi_1}\bra{\psi_2'}+\ket{\psi_1'}\bra{\psi_2}+\ket{\psi_2}\bra{\psi_1'}+\\
\nonumber &+\ket{\psi_2'}\bra{\psi_1})-\frac{2}{3}(\ket{\psi_2}\bra{\psi_2'}+\ket{\psi_2'}\bra{\psi_2})\;\;\;j=1,2\\
\nonumber A_3(\omega_8)=&-(\ket{\psi_1}\bra{\psi_1'}+\ket{\psi_1'}\bra{\psi_1})+\frac{1}{3}(\ket{\psi_2}\bra{\psi_2'}+\\
\nonumber &\ket{\psi_2'}\bra{\psi_2}).
\end{eqnarray}

\section{Appendix C}
As far as the third step it is useful to rewrite the jump
operators of the first step in the basis $\ket{\widetilde{000}}$,
$\ket{\widetilde{111}}$, $\ket{\widetilde{W}}$,
$\ket{\widetilde{W'}}$,
$\ket{\widetilde{\psi_1}}$, $\ket{\widetilde{\psi_1'}}$, $\ket{\widetilde{\psi_2}}$ and $\ket{\widetilde{\psi_2'}}$\\

\begin{eqnarray}
\nonumber A_j(\omega_1)=&-\frac{1}{2\sqrt{3}}\bigg[\big(\cos{\beta}+i\sqrt{2}\sin{\beta}\big)\big(\sqrt{2}\ket{\widetilde{W}}\bra{\widetilde{000}}+\\
\nonumber &+\sqrt{3}\ket{\widetilde{\psi_1}}\bra{\widetilde{000}}+\ket{\widetilde{\psi_2}}\bra{\widetilde{000}}\big)
+\big(\cos{\beta}-i\sqrt{2}\sin{\beta}\big)\times\\
\nonumber &\times\big(\sqrt{2}\ket{\widetilde{W'}}\bra{\widetilde{111}}+\sqrt{3}\ket{\widetilde{\psi_1'}}\bra{\widetilde{111}}
+\ket{\widetilde{\psi_2'}}\bra{\widetilde{111}}\big)
\bigg]\;\;\;j=1,2\\
\nonumber A_3(\omega_1)=&-\frac{1}{\sqrt{6}}\bigg[\big(\cos{\beta}+i\sqrt{2}\sin{\beta}\big)\big(\ket{\widetilde{W}}\bra{\widetilde{000}}+\\
\nonumber &-\sqrt{2}\ket{\widetilde{\psi_2}}\bra{\widetilde{000}}\big)+\big(\cos{\beta}-i\sqrt{2}\sin{\beta}\big)\big(\ket{\widetilde{W'}}\bra{\widetilde{111}}+\\
&-\sqrt{2}\ket{\widetilde{\psi_2'}}\bra{\widetilde{111}}\big)\bigg]\\
\nonumber A_j(\omega_2)=&-\frac{\cos{\beta}}{3\sqrt{2}}\bigg[3\ket{\widetilde{000}}\bra{\widetilde{000}}-3\ket{\widetilde{111}}\bra{\widetilde{111}}+
\ket{\widetilde{W}}\bra{\widetilde{W}}+\\&-\ket{\widetilde{W'}}\bra{\widetilde{W'}}+2\ket{\widetilde{\psi_2}}\bra{\widetilde{\psi_2}}-
\nonumber 2\ket{\widetilde{\psi_2'}}\bra{\widetilde{\psi_2'}}\bigg]\;\;\;j=1,2\\
\nonumber A_3(\omega_2)=&-\frac{\cos{\beta}}{3\sqrt{2}}\bigg[3\ket{\widetilde{000}}\bra{\widetilde{000}}-3\ket{\widetilde{111}}\bra{\widetilde{111}}+\\
\nonumber &+\ket{\widetilde{W}}\bra{\widetilde{W}}-\ket{\widetilde{W'}}\bra{\widetilde{W'}}+3\ket{\widetilde{\psi_1}}\bra{\widetilde{\psi_1}}+\\
\nonumber &-3\ket{\widetilde{\psi_1'}}\bra{\widetilde{\psi_1'}}+2\ket{\widetilde{\psi_2}}\bra{\widetilde{\psi_2}}-2\ket{\widetilde{\psi_2'}}
\bra{\widetilde{\psi_2'}}\bigg].
\end{eqnarray}

\end{document}